\documentclass[prb,superscriptaddress,showpacs,byrevtex,twocolumn]{revtex4}
\usepackage[]{amssymb}
\usepackage[]{graphicx}

\begin{document}

\bibliographystyle{apsrev}

\title{Echo of the Quantum Phase Transition of CeCu$_{6-x}$Au$_x$ in XPS:\\
Breakdown of Kondo Screening}

\author{M. Klein}
\affiliation{Universit\"at W\"urzburg, Experimentelle Physik II,
  Am Hubland, D-97074 W\"urzburg, Germany} 
\author{J. Kroha}
\affiliation{Physikalisches Institut and Bethe Center for Theoretical 
Physics, Universit\"at Bonn, 
  Nussallee 12, D-53115 Bonn, Germany} 
\author{H. v.\ L\"ohneysen}
\affiliation{Physikalisches Institut,
Universit\"at Karlsruhe (TH), D-76128 Karlsruhe, Germany}
\affiliation{Forschungszentrum Karlsruhe, Institut f\"ur Festk\"operphysik, 
D-76021 Karlsruhe, Germany}
\author{O. Stockert}
\affiliation{Max Planck Institute for Chemical Physics of Solids, 
N\"othnitzer Str. 40, 01187 Dresden, Germany}
\author{F. Reinert}
\affiliation{Universit\"at W\"urzburg, Experimentelle Physik II,
  Am Hubland, D-97074 W\"urzburg, Germany} 
\affiliation{Forschungszentrum Karlsruhe, Gemeinschaftslabor f\"ur Nanoanalytik, D-76021 Karlsruhe, Germany}

\date{\today}

\begin{abstract}
We present an X-ray photoemission study of the heavy fermion system
CeCu$_{6-x}$Au$_x$ across the magnetic quantum phase transition 
of this compound at temperatures above the single-ion Kondo temperature $T_K$.
In dependence of the Au concentration $x$ we observe a sudden change of the 
$f$-occupation number $n_f$ and the core-hole potential $U_{df}$ at 
the critical concentration $x_c=0.1$. We interpret these findings in the 
framework of the single-impurity Anderson model. Our results are in excellent 
agreement with findings from earlier UPS measurements 
and provide further information about the precursors of quantum criticality at 
elevated temperatures.
\end{abstract}
\maketitle

\section{Introduction}
CeCu$_{6-x}$Au$_x$ is one of the best characterized heavy-fermion (HF) 
compounds\cite{loehneysen00, stockert98,schroeder00, loehneysen98a, 
loehneysen06} exhibiting a quantum phase transition (QPT) between a paramagnetic 
phase with full Kondo screening of the Ce 4f moments and an antiferromagnetic 
phase induced by their RKKY coupling. \cite{gegenwart08} 
The QPT occurs at a critical Au concentration of $x=x_c = 0.1$. 
Two major scenarios for the QPT have been put forward, firstly the
Hertz-Millis scenario\cite{hertz76,millis93}  where only the (bosonic) 
magnetic fluctuations become critical, leaving the (fermionic) heavy 
quasiparticles intact, and secondly the local quantum 
criticality\cite{qsi01,coleman01} where the local Kondo quasiparticles are 
destroyed by coupling to the quantum critical magnetic fluctuations. 
In the latter case the Kondo screening scale, or Kondo temperature of the 
lattice system, $T_K^{*}$, vanishes at the quantum critical point (QCP). 
Despite intense efforts it has not been possible to unambiguously identify 
either one scenario in CeCu$_{6-x}$Au$_x$. The difficulty resides in the 
fact that near the QCP many different effects come into play, including 
local Kondo screening, lattice coherence and Fermi volume collapse, 
quantum critical fluctuations, and possible dimensional 
reduction.\cite{stockert98,rosch97} They have prevented a conclusive 
experimental picture as well as a unified theory of the QPT. Recently we
have presented direct measurements of the evolution of the local Kondo 
screening energy scale $T_K$ across the QCP as extracted from  
high-resolution 
ultraviolet photoemission spectroscopy (UPS).\cite{klein08qpt}  
Taking impurity spectra clearly above $T_K$ as well as above the lattice 
coherence temperature $T_{coh}$ and the N\'eel temperature for magnetic 
ordering, $T_N$ ( $T>\{T_K, T_{coh}, T_N\}$), made it possible to probe the 
local energy scale without the complications caused by lattice coherence 
or quantum critical fluctuations. The surprising 
outcome of these investigations is that the precursors 
of the QPT are already visible in the single-impurity $T_K$ as extracted
from UPS spectra at such elevated temperatures.
In Ref.~[\onlinecite{klein08qpt}] we developed a theory, based on the 
effective single-impurity model realized for temperatures $T>T_{coh}$, 
where the Kondo 
exchange coupling $J$ of a Ce site is selfconsistently renormalized by the 
surrounding, identical Ce atoms only through the indirect RKKY coupling. 
The theory explains the observed, nontrivial, steplike behavior of $T_K$ 
near the QCP and supports the local critical scenario for CeCu$_{6-x}$Au$_x$. 
It also provides a general criterion to distinguish experimentally 
whether a given HF compound should exhibit HM or local quantum critical 
behaviour.\cite{klein08qpt}

UPS provides the most direct access
to the screening scale $T_K$ by directly recording the Kondo resonance in the
local Ce $4f$ spectrum,\cite{ehm07, allen05} but probes a relatively
shallow surface region only. Therefore we complement here the studies 
of Ref.~[\onlinecite{klein08qpt}] with 
bulk-sensitive X-ray photoemission spectroscopy (XPS) on the same samples
and under the same experimental conditions, even though XPS provides only 
rather indirect information about the Ce $4f$ spectrum and the physics 
near the Fermi energy $E_F$.\cite{reinert01cr,schmidt05}
We find that the XPS results are in full agreement with the UPS analysis.

\section{Significance of core-hole spectra}
Before we show the experimental results, 
we briefly recall the energetics of the photoionization process, in order
to understand the essential structure of the XPS spectra. 
In XPS the Ce $3d$ spectrum is recorded in the final state of the Ce ion after 
creation of a $3d$ core hole. The binding energy of a $3d$ electron is 
influenced by the electrostatic attraction $-U_{df}$ between the $3d$ core
hole and a $4f$ electron and, hence, depends on the number of electrons $n$ in 
the $4f$ shell. Therefore, the $3d$ peak of the XPS spectrum is split into 
individual resonances, each one corresponding to a different charge state 
of the $4f$ shell, the so-called $f^n$ resonances, with $n=0,~1,~2$.  
In addition, there is a Coulomb repulsion $U_{ff}$ between two electrons 
in the $4f$ shell. The peak position (binding energy $E_B^{(n)}$) of 
the $f^n$ resonance is the energy difference between the charge 
configuration in the photoionized final state and in the groundstate. 
Hence, it is roughly given by, up to conduction electron screening corrections,
\begin{eqnarray}
E_B^{(n)} = E_B^{(0)} - n\, U_{df} + \frac{1}{2} (n^2-n)\, U_{ff} \ .
\label{eq:EB}
\end{eqnarray}
Note that the attractive core hole potential $-U_{df}$
changes the whole level scheme.\cite{huefner} In particular, in the 
photoionized state the $4f^2$ configuration becomes occupied 
(see Fig.~\ref{fig:stack}), even though in the Ce groundstate it is 
shifted far above the Fermi energy by the repulsion $U_{ff}$. The  
quantitative fit of $U_{df}$ and $U_{ff}$ (see below) shows 
that the binding energy of the $4f^0$ configuration, $E_B^{(0)}$, 
is the largest. The spectral weight of the $f^0$ XPS peak is  
proportional to the Kondo resonance of the Ce $4f$ spectrum, 
since in CeCu$_{6-x}$Au$_{x}$ most of the Kondo peak weight is located above 
$E_F$\cite{ehm07} and thus represents the unoccupied configuration of 
the $4f$ shell. 

\section{Experimental results}
All the measurements were carried out at $T=15$~K and thus at temperatures 
far above any long-range or quantum critical fluctuations. The experimental 
setup is equipped with a SCIENTA R4000 analyser and a monochromatized X-ray 
source (Al K$_\alpha$ line: X-ray enery $h\nu = 1486$~eV). 
The pressure in the UHV chamber 
was $1 \times 10^{-10}$~mbar. Because of the low intensity due to the small 
surface area of the crytals and the rapid detoriation of the surface, the 
instrumental resolution was set to $\approx 2$~eV in order to optimize 
the count rate. The investigated samples were single crystals grown in a W 
crucible under high-purity Ar atmosphere \cite{grube99} and 
cleaved \textit{in situ} at the sample position just before the measurement.
   
\begin{figure}[t]
 \begin{center}
  \includegraphics[width=0.95\linewidth]{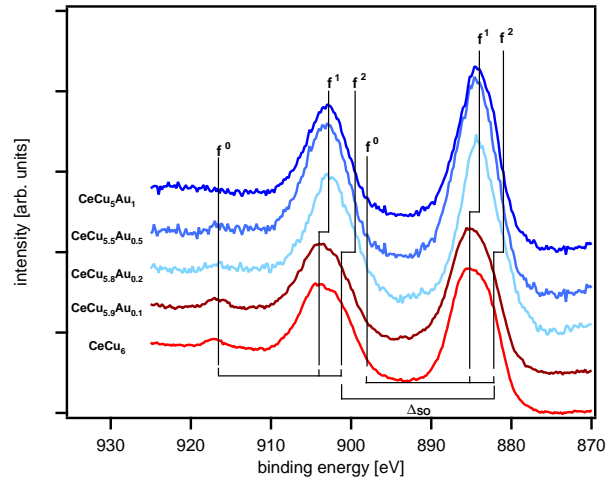}
  \caption[he1]{(color online) $3d$ core-level spectra of CeCu$_{6-x}$Au$_x$ 
  for five
  differnet Au concentrations at $T=15$~K (X-ray energy $h\nu=1486$~eV,
  experimental resolution $\Delta E\approx2$~eV). The vertical lines mark
the peak positions exhibiting an abrupt shift of the $f^1$ and $f^2$ 
resonances from $x=0.1$ to $x=0.2$. 
  \label{fig:stack}}
 \end{center}
\end{figure}

The Ce $3d$ core-level spectra for all measured compounds are shown 
in Fig.~\ref{fig:stack}. 
Taking the second derivative of each spectrum reveals that each of 
the broad peaks is actually comprised of two resonances. Hence, each spectrum consists of 
a group of three resonances, termed $f^0$, $f^1$, and $f^2$ in 
Fig.~\ref{fig:stack}. They are, in addition, duplicated due to a
spin-orbit splitting of $\Delta_{\text{SO}} = 18.9$~eV.
A rough analysis already shows that the XPS $3d$ spectra can be  
divided into two classes, namely for $x\le 0.1$ and $x\ge 0.2$, respectively,
like in the case of the UPS $4f$ spectra: \cite{klein08qpt}
For $x\le 0.1$ the $f^0$ peak at $E_B\approx 917$~eV is clearly visible, 
albeit comparatively weak, while for $x\ge 0.2$ the $f^0$ weight is 
almost not discernible and vanishes completely for $x=1$. 
In addition, for $x\ge 0.2$
the $f^1$ and $f^2$ peaks are shifted by a constant amount of 1.8~eV 
towards smaller binding energy relative to their positions for $x\le 0.1$.
\begin{figure}[b]
 \begin{center}
  \includegraphics[width=0.95\linewidth]{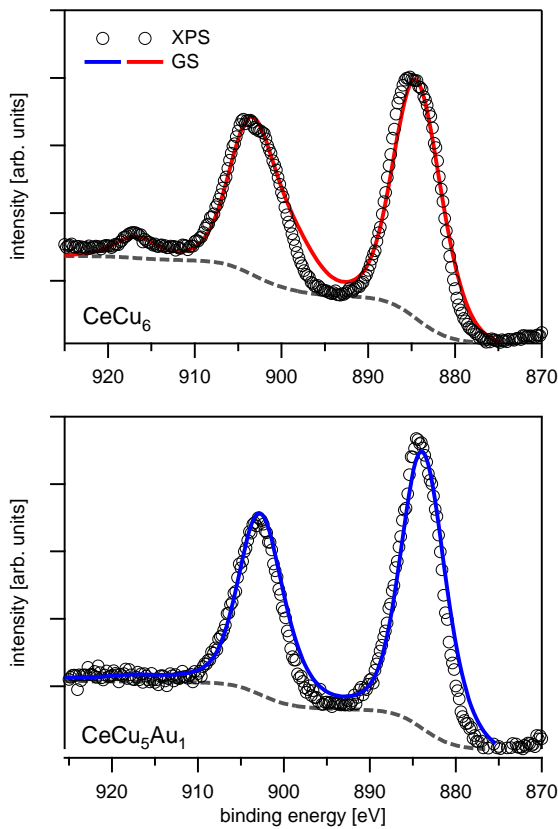}
  \caption[he1]{(color online) Comparison of the experimental XPS data and the theoretical spectra calculated from the GS theory.\cite{gunnarsson83,gunnarsson83b} The dashed lines represent the inelastic background that was added phenomenologically to the theoretical spectra. 
  \label{fig:vgl}}
 \end{center}
\end{figure}

\section{Discussion}
\begin{figure}[tb]
 \begin{center}
  \includegraphics[width=0.95\linewidth]{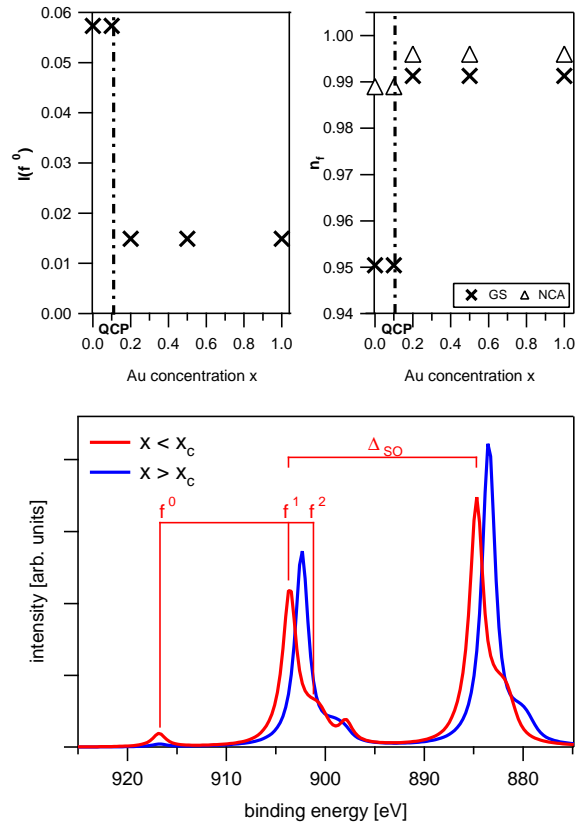}
  \caption[he1]{(color online) The upper left panel shows the weight 
of the $f^0$-configuration as determined from the GS fit. It 
represents a measure of the weight of the Kondo-resonance
in the $4f$ spectrum (see text). 
The upper right panel shows the $4f$-occupation numbers. In the lower 
panel the calculated, unbroadened 
GS spectra are shown 
for CeCu$_{6-x}$Au$_{x}$ compounds below and above the quantum critical 
concentration $x_c$. 
The parameter values used in the calculations are given in the text. 
The vertical lines mark the $f^0$, $f^1$, $f^2$ peaks and the 
SO-shifting for $x<x_c$. 
  \label{fig:nf}}
 \end{center}
\end{figure}

In order to interpret these XPS results quantitatively, we
calculated the Ce $3d$ spectra using the Gunnarson--Sch\"onhammer 
(GS) theory,\cite{gunnarsson83,gunnarsson83b} which is based on the 
single-impurity Anderson model (SIAM). From earlier UPS experiments
it is known\cite{klein08qpt,ehm01,ehm07} that the 
CeCu$_{6-x}$Au$_{x}$ $4f$ spectra and their $T$ dependence 
are very well described in the framework of this model for elevated 
temperatures $T>T_K$, with an RKKY-induced renormalization 
of the local Kondo coupling.\cite{klein08qpt} Moreover, the GS theory 
has been successfully applied to interpret XPS, X-ray photoabsorption, 
inverse photoemission and valence photoemission spectroscopy on other Ce-based 
compounds, using the same parameter set for all 
measurements,\cite{fuggle83,allen86} albeit
small adjustments of parameter values were pointed out to be necessary 
in some cases.\cite{witkowski97b, witkowski01} In our GS calculations 
for CeCu$_{6-x}$Au$_{x}$ we use, without fitting, the same 
parameter values of the SIAM that were determined earlier from  
noncrossing approximation (NCA) calculations of the UPS $4f$ spectra 
of the same 
compounds.\cite{klein08qpt} That is, the conduction band half width and the
bare $4f$ level are $D=2.8$~eV and $\varepsilon_f=-1.05$~eV, respectively.
The $4f$-level hybridization had been determined as 
$\Delta = 116$~meV for $x\leq 0.1$ 
and $\Delta = 108$~meV for $x\geq 0.2$, corresponging to a 
reduction of the effective spin exchange coupling 
$J= \Delta^2[1/(-\varepsilon_f) + 1/(\varepsilon_f+U_{ff})]$. 
The XPS $3d$ spectra involve the
additional interaction parameters $U_{df}$ and $U_{ff}$, see above.
Since the intra-$4f$-level repulsion $U_{ff}$ will not significantly
depend on the Au concentration $x$, we have chosen a constant value of
$U_{ff}=10.0$~eV~$\gg |\epsilon_f|$ (within NCA it was assumed to be infinite
for simplicity, in order to suppress $4f$ double occupancy). 
This leaves the core hole potential $U_{df}$ as the only adjustable parameter.
It determines the final-state energy levels and thus the position of the 
core-level peaks relative to each other according to Eq.~(\ref{eq:EB}). 
To fit the theoretical spectra to the experimental results we added to 
the GS spectra a background, accounting phenomenologically for inelastic 
scattering of the electrons during the
XPS process, and convoluted the theoretical spectra with a Gaussian, 
accounting for the experimental resolution. The fits yield  
$U_{df}=12.5$~eV for $x\le 0.1$ and $U_{df}=13.5$~eV for $x\ge 0.2$.
Fig.~\ref{fig:vgl} shows the GS spectra, fitted to the experimental data, and the lower panel of Fig.~\ref{fig:nf} shows a comparisson of the unbroadened GS spectra below and above $x_c$. 
The agreement between theory and experiment is remarkable: 
Both, the relative peak intensities and the shape of the spectra are 
well reproduced by only a single fit parameter. 

This allows for a detailed interpretation of the experimental data.
A reduction of $\Delta$ leads to a significant change in the relative 
weights of the peaks, as already shown by Gunnarsson and Sch\"onhammer in 
Refs.~[\onlinecite{gunnarsson83, gunnarsson83b}]: For $\Delta=108$~meV
the $f^2$-peak is diminished and the $f^0$-peak has almost vanished. 
Since the weight of the $f^0$-configuration is a measure of the
weigth of the Kondo resonance (see above), this means for the 
present CeCu$_{6-x}$Au$_{x}$ system that 
the spectral weight of the Kondo resonance drops abruptly as $x$ is 
changed from $x\le x_c$ to $x>x_c$, in full agreement with the findings
of UPS in Ref.~[\onlinecite{klein08qpt}]. In the 
left upper panel of Fig.~\ref{fig:nf} we show the 
weight of the $f^0$-configuration as extracted from 
the GS calculations for different $x$. It shows a significant step at $x=x_c$.
As a consequence of this collapse 
of the Kondo resonance in the $4f$ spectrum above the Fermi level, 
one expects that the spectral weight below $E_F$ and, hence, 
the $f$-level occupation number $n_f$ are increased for $x>x_c$
by roughly the Kondo spectral weight.
As an immediate, further consequence the $3d$ electron binding energy 
should decrease, because the enhancement of $n_f$ increases the 
attractive interaction exerted by the $4f$ electrons on the $3d$ core hole.

Both expectations can be checked from the XPS spectra. Firstly,
the decrease of $3d$ binding energy is indeed seen 
in the experimental spectra (Fig.~\ref{fig:stack}) and in the fitted
theoretical spectra (Fig.~\ref{fig:nf}, lower panel) as the upward shift of 
the $f^1$ and $f^2$ peaks for $x\geq 0.2$ as compared to $x\leq 0.1$. It is
also reflected by the increase of the fitted values of the
 effective core level potential $U_{df}$. 
Secondly, the probability for the $4f$ occupancy in the 
groundstate can be extracted from the XPS spectra.\cite{reinert01cr,schmidt05}
It is roughly given by
\begin{eqnarray} 
n_f=1-\frac{I(f^0)}{I(f^0)+I(f^1)+2\, I(f^2)} \ ,
\label{eq:nf}
\end{eqnarray} 
where $I(f^n)$ is the (unnormalized) spectral weight of the $f^n$ 
peak, $n=0,~1,~2$,
after substraction of the inelastic background. Determining $n_f$
directly from the experimental data is not accurate, since 
the $f^n$-peaks overlap and the determination of the inelastic
background of the experimental spectra is not trivial. 
Instead we extract the $n_f$-numbers from 
the fitted GS calculations. These include also
small deviations from Eq. (\ref{eq:nf}) which arise from the dynamics
of the XPS process.\cite{fuggle83,gunnarsson83b} 
In the right upper panel of Fig.~\ref{fig:nf} we compare the $n_f$ values 
obtained from the GS calculations and from NCA calculations 
of Ref.~\onlinecite{klein08qpt}. 
Since $n_f$ increases with temperature,\cite{bickers87} it is 
important to note that the NCA values have been calculated at the experimental 
temperature of $T=15$~K. Although the absolute values of $n_f$ differ for the
two calculations -- probably due to the use of a infinite $U_{ff}$ within NCA --
the main features are the same: both data sets exhibit a significant, 
abrupt step of $n_f$ at the critical concentration $x_c$ for
the QPT in CeCu$_{6-x}$Au$_{x}$, as expected from the Kondo resonance
collapse observed for $x>0.1$.

\section{Conclusion}
We have presented data from XPS-measurements at elevated 
temperature on CeCu$_{6-x}$Au$_x$ compounds with five different Au 
concentrations $x$ across the critical concentration $x_c=0.1$ of the magnetic
quantum phase transition. With increasing $x$ the $3d$ core spectra exhibit 
near $x=x_c$ an abrupt change with respect to three different features, namely
(1) a collapse of the $f^0$ resonance, signalling a sudden decrease of Kondo
screening, (2) a step-like increase of the $4f$ occupation number and 
(3) a sudden increase of the core hole attraction, evidenced by a 
shift of the $f^1$ and $f^2$ peaks towards smaller binding energy.
The first feature is in complete agreement with findings of direct UPS
measurements of the Kondo resonance in the Ce $4f$ spectrum. The features
(2) and (3) are explained in a natural way as direct consequences of the
Kondo resonance collapse. Therefore, the present measurements strongly support
that the Kondo resonance collapse observed at the QPT by UPS \cite{klein08qpt} 
is not a surface effect but an effect of the bulk CeCu$_{6-x}$Au$_x$ 
compounds, since it is seen by the bulk-sensitive XPS as well.  
The analysis of the spectra using the Gunnarson-Sch\"onhammer 
theory\cite{gunnarsson83,gunnarsson83b}, supported by NCA calculations,
further indicates that the spectra taken at temperatures above
the single-impurity Kondo temperature $T_K$ and above the lattice 
coherence and N\'eel temperatures $T_{coh}$, $T_{N}$ are well described
by the single-impurity Anderson model, with the spin exchange coupling $J$
renormalized only by RKKY interactions with the surrounding Ce impurity 
moments.\cite{klein08qpt}
Thus, our results indicate experimentally that the HM 
model\cite{hertz76,millis93} is
insufficient to describe the QPT in CeCu$_{6-x}$Au$_{x}$ and that at the
lowest temperatures the system should follow the local quantum critical 
scenario,\cite{qsi01,coleman01} as argued in detail 
in Ref.~\onlinecite{klein08qpt}.

We would like to thank O. Gunnarsson for fruitful disscussions and
for the allocation of his code.
This work was supported by the Deutsche Forschungsgemeinschaft  through 
grant No. Re~1469/4-3/4 (M.K., F.R.), FOR 960 (H.v.L.), and SFB 608
(J.K.).


\begin{thebibliography}{26}
\expandafter\ifx\csname natexlab\endcsname\relax\def\natexlab#1{#1}\fi
\expandafter\ifx\csname bibnamefont\endcsname\relax
  \def\bibnamefont#1{#1}\fi
\expandafter\ifx\csname bibfnamefont\endcsname\relax
  \def\bibfnamefont#1{#1}\fi
\expandafter\ifx\csname citenamefont\endcsname\relax
  \def\citenamefont#1{#1}\fi
\expandafter\ifx\csname url\endcsname\relax
  \def\url#1{\texttt{#1}}\fi
\expandafter\ifx\csname urlprefix\endcsname\relax\def\urlprefix{URL }\fi
\providecommand{\bibinfo}[2]{#2}
\providecommand{\eprint}[2][]{\url{#2}}

\bibitem[{\citenamefont{v.~L\"ohneysen
  et~al.}(2000)\citenamefont{v.~L\"ohneysen, Schr\"oder, and
  Stockert}}]{loehneysen00}
\bibinfo{author}{\bibfnamefont{H.}~\bibnamefont{v.~L\"ohneysen}},
  \bibinfo{author}{\bibfnamefont{A.}~\bibnamefont{Schr\"oder}},
  \bibnamefont{and} \bibinfo{author}{\bibfnamefont{O.}~\bibnamefont{Stockert}},
  \bibinfo{journal}{J. Alloys Compd.} \textbf{\bibinfo{volume}{303--304}},
  \bibinfo{pages}{480} (\bibinfo{year}{2000}).

\bibitem[{\citenamefont{Stockert et~al.}(1998)\citenamefont{Stockert,
  v.~L\"ohneysen, Rosch, Pyka, and Loewenhaupt}}]{stockert98}
\bibinfo{author}{\bibfnamefont{O.}~\bibnamefont{Stockert}},
  \bibinfo{author}{\bibfnamefont{H.}~\bibnamefont{v.~L\"ohneysen}},
  \bibinfo{author}{\bibfnamefont{A.}~\bibnamefont{Rosch}},
  \bibinfo{author}{\bibfnamefont{N.}~\bibnamefont{Pyka}}, \bibnamefont{and}
  \bibinfo{author}{\bibfnamefont{M.}~\bibnamefont{Loewenhaupt}},
  \bibinfo{journal}{Phys. Rev. Lett.} \textbf{\bibinfo{volume}{80}},
  \bibinfo{pages}{5627} (\bibinfo{year}{1998}).

\bibitem[{\citenamefont{Schr\"oder et~al.}(2000)\citenamefont{Schr\"oder,
  Aeppli, Coldea, Adams, Stockert, v.~L\"{o}hneysen, Bucher, Ramazashvili, and
  Coleman}}]{schroeder00}
\bibinfo{author}{\bibfnamefont{A.}~\bibnamefont{Schr\"oder}},
  \bibinfo{author}{\bibfnamefont{G.}~\bibnamefont{Aeppli}},
  \bibinfo{author}{\bibfnamefont{R.}~\bibnamefont{Coldea}},
  \bibinfo{author}{\bibfnamefont{M.}~\bibnamefont{Adams}},
  \bibinfo{author}{\bibfnamefont{O.}~\bibnamefont{Stockert}},
  \bibinfo{author}{\bibfnamefont{H.}~\bibnamefont{v.~L\"{o}hneysen}},
  \bibinfo{author}{\bibfnamefont{E.}~\bibnamefont{Bucher}},
  \bibinfo{author}{\bibfnamefont{R.}~\bibnamefont{Ramazashvili}},
  \bibnamefont{and} \bibinfo{author}{\bibfnamefont{P.}~\bibnamefont{Coleman}},
  \bibinfo{journal}{Nature} \textbf{\bibinfo{volume}{407}},
  \bibinfo{pages}{351} (\bibinfo{year}{2000}).

\bibitem[{\citenamefont{v.~L\"ohneysen
  et~al.}(1998)\citenamefont{v.~L\"ohneysen, Neubert, Pietrus, Schr\"oder,
  Stockert, Tutsch, L\"owenhaupt, Rosch, and W\"olfle}}]{loehneysen98a}
\bibinfo{author}{\bibfnamefont{H.}~\bibnamefont{v.~L\"ohneysen}},
  \bibinfo{author}{\bibfnamefont{A.}~\bibnamefont{Neubert}},
  \bibinfo{author}{\bibfnamefont{T.}~\bibnamefont{Pietrus}},
  \bibinfo{author}{\bibfnamefont{A.}~\bibnamefont{Schr\"oder}},
  \bibinfo{author}{\bibfnamefont{O.}~\bibnamefont{Stockert}},
  \bibinfo{author}{\bibfnamefont{U.}~\bibnamefont{Tutsch}},
  \bibinfo{author}{\bibfnamefont{M.}~\bibnamefont{L\"owenhaupt}},
  \bibinfo{author}{\bibfnamefont{A.}~\bibnamefont{Rosch}}, \bibnamefont{and}
  \bibinfo{author}{\bibfnamefont{P.}~\bibnamefont{W\"olfle}},
  \bibinfo{journal}{Eur. Phys. J. B} \textbf{\bibinfo{volume}{5}},
  \bibinfo{pages}{447} (\bibinfo{year}{1998}).

\bibitem[{\citenamefont{v.~L{\"o}hneysen
  et~al.}(2006)\citenamefont{v.~L{\"o}hneysen, Bartolf, Drotziger, Pfleiderer,
  Stockert, Souptel, L{\"o}ser, and Behr}}]{loehneysen06}
\bibinfo{author}{\bibfnamefont{H.}~\bibnamefont{v.~L{\"o}hneysen}},
  \bibinfo{author}{\bibfnamefont{H.}~\bibnamefont{Bartolf}},
  \bibinfo{author}{\bibfnamefont{S.}~\bibnamefont{Drotziger}},
  \bibinfo{author}{\bibfnamefont{C.}~\bibnamefont{Pfleiderer}},
  \bibinfo{author}{\bibfnamefont{O.}~\bibnamefont{Stockert}},
  \bibinfo{author}{\bibfnamefont{D.}~\bibnamefont{Souptel}},
  \bibinfo{author}{\bibfnamefont{W.}~\bibnamefont{L{\"o}ser}},
  \bibnamefont{and} \bibinfo{author}{\bibfnamefont{G.}~\bibnamefont{Behr}},
  \bibinfo{journal}{J. Alloys Compd.} \textbf{\bibinfo{volume}{408--412}},
  \bibinfo{pages}{9} (\bibinfo{year}{2006}).

\bibitem[{\citenamefont{Gegenwart et~al.}(2008)\citenamefont{Gegenwart, Si, and
  Steglich}}]{gegenwart08}
\bibinfo{author}{\bibfnamefont{P.}~\bibnamefont{Gegenwart}},
  \bibinfo{author}{\bibfnamefont{Q.}~\bibnamefont{Si}}, \bibnamefont{and}
  \bibinfo{author}{\bibfnamefont{F.}~\bibnamefont{Steglich}},
  \bibinfo{journal}{Nature Phys.} \textbf{\bibinfo{volume}{4}},
  \bibinfo{pages}{186} (\bibinfo{year}{2008}).

\bibitem[{\citenamefont{Hertz}(1976)}]{hertz76}
\bibinfo{author}{\bibfnamefont{J.~A.} \bibnamefont{Hertz}},
  \bibinfo{journal}{Phys. Rev. B} \textbf{\bibinfo{volume}{14}},
  \bibinfo{pages}{1165} (\bibinfo{year}{1976}).

\bibitem[{\citenamefont{Millis}(1993)}]{millis93}
\bibinfo{author}{\bibfnamefont{A.~J.} \bibnamefont{Millis}},
  \bibinfo{journal}{Phys. Rev. B} \textbf{\bibinfo{volume}{48}},
  \bibinfo{pages}{7183} (\bibinfo{year}{1993}).

\bibitem[{\citenamefont{Si et~al.}(2001)\citenamefont{Si, Rabello, and
  Smith}}]{qsi01}
\bibinfo{author}{\bibfnamefont{Q.}~\bibnamefont{Si}},
  \bibinfo{author}{\bibfnamefont{S.}~\bibnamefont{Rabello}}, \bibnamefont{and}
  \bibinfo{author}{\bibfnamefont{J.~L.} \bibnamefont{Smith}},
  \bibinfo{journal}{Nature} \textbf{\bibinfo{volume}{413}},
  \bibinfo{pages}{804} (\bibinfo{year}{2001}).

\bibitem[{\citenamefont{Coleman et~al.}(2001)\citenamefont{Coleman, P\'epin,
  Si, and Ramazashvili}}]{coleman01}
\bibinfo{author}{\bibfnamefont{P.}~\bibnamefont{Coleman}},
  \bibinfo{author}{\bibfnamefont{C.}~\bibnamefont{P\'epin}},
  \bibinfo{author}{\bibfnamefont{Q.}~\bibnamefont{Si}}, \bibnamefont{and}
  \bibinfo{author}{\bibfnamefont{R.}~\bibnamefont{Ramazashvili}},
  \bibinfo{journal}{J. Phys.: Condens. Matter} \textbf{\bibinfo{volume}{13}},
  \bibinfo{pages}{723} (\bibinfo{year}{2001}).

\bibitem[{\citenamefont{Rosch et~al.}(1997)\citenamefont{Rosch, Schr\"oder,
  Stockert, and v.~L\"ohneysen}}]{rosch97}
\bibinfo{author}{\bibfnamefont{A.}~\bibnamefont{Rosch}},
  \bibinfo{author}{\bibfnamefont{A.}~\bibnamefont{Schr\"oder}},
  \bibinfo{author}{\bibfnamefont{O.}~\bibnamefont{Stockert}}, \bibnamefont{and}
  \bibinfo{author}{\bibfnamefont{H.}~\bibnamefont{v.~L\"ohneysen}},
  \bibinfo{journal}{Phys. Rev. Lett.} \textbf{\bibinfo{volume}{79}},
  \bibinfo{pages}{159} (\bibinfo{year}{1997}).

\bibitem[{\citenamefont{Klein et~al.}(2008)\citenamefont{Klein, Nuber, Reinert,
  Kroha, and v.~L\"ohneysen}}]{klein08qpt}
\bibinfo{author}{\bibfnamefont{M.}~\bibnamefont{Klein}},
  \bibinfo{author}{\bibfnamefont{A.}~\bibnamefont{Nuber}},
  \bibinfo{author}{\bibfnamefont{F.}~\bibnamefont{Reinert}},
  \bibinfo{author}{\bibfnamefont{J.}~\bibnamefont{Kroha}}, \bibnamefont{and}
  \bibinfo{author}{\bibfnamefont{H.}~\bibnamefont{v.L\"ohneysen}},
  \bibinfo{journal}{Phys. Rev. Lett.} \textbf{\bibinfo{volume}{101}},
  \bibinfo{pages}{266404} (\bibinfo{year}{2008}).

\bibitem[{\citenamefont{Ehm et~al.}(2007)\citenamefont{Ehm, H\"ufner, Reinert,
  Kroha, W\"olfle, Stockert, Geibel, and v.~L\"ohneysen}}]{ehm07}
\bibinfo{author}{\bibfnamefont{D.}~\bibnamefont{Ehm}},
  \bibinfo{author}{\bibfnamefont{S.}~\bibnamefont{H\"ufner}},
  \bibinfo{author}{\bibfnamefont{F.}~\bibnamefont{Reinert}},
  \bibinfo{author}{\bibfnamefont{J.}~\bibnamefont{Kroha}},
  \bibinfo{author}{\bibfnamefont{P.}~\bibnamefont{W\"olfle}},
  \bibinfo{author}{\bibfnamefont{O.}~\bibnamefont{Stockert}},
  \bibinfo{author}{\bibfnamefont{C.}~\bibnamefont{Geibel}}, \bibnamefont{and}
  \bibinfo{author}{\bibfnamefont{H.}~\bibnamefont{v.~L\"ohneysen}},
  \bibinfo{journal}{Phys. Rev. B} \textbf{\bibinfo{volume}{76}},
  \bibinfo{pages}{045117} (\bibinfo{year}{2007}).

\bibitem[{\citenamefont{Allen}(2005)}]{allen05}
\bibinfo{author}{\bibfnamefont{J.~W.} \bibnamefont{Allen}},
  \bibinfo{journal}{J. Phys. Soc. Jpn.} \textbf{\bibinfo{volume}{74}},
  \bibinfo{pages}{34} (\bibinfo{year}{2005}).

\bibitem[{\citenamefont{Reinert et~al.}(2001)\citenamefont{Reinert, Claessen,
  Nicolay, Ehm, H\"ufner, Ellis, Gweon, Allen, Kindler, and
  A\ss{}mus}}]{reinert01cr}
\bibinfo{author}{\bibfnamefont{F.}~\bibnamefont{Reinert}},
  \bibinfo{author}{\bibfnamefont{R.}~\bibnamefont{Claessen}},
  \bibinfo{author}{\bibfnamefont{G.}~\bibnamefont{Nicolay}},
  \bibinfo{author}{\bibfnamefont{D.}~\bibnamefont{Ehm}},
  \bibinfo{author}{\bibfnamefont{S.}~\bibnamefont{H\"ufner}},
  \bibinfo{author}{\bibfnamefont{W.~P.} \bibnamefont{Ellis}},
  \bibinfo{author}{\bibfnamefont{G.~H.} \bibnamefont{Gweon}},
  \bibinfo{author}{\bibfnamefont{J.~W.} \bibnamefont{Allen}},
  \bibinfo{author}{\bibfnamefont{B.}~\bibnamefont{Kindler}}, \bibnamefont{and}
  \bibinfo{author}{\bibfnamefont{W.}~\bibnamefont{A\ss{}mus}},
  \bibinfo{journal}{Phys. Rev. B} \textbf{\bibinfo{volume}{63}},
  \bibinfo{pages}{197102} (\bibinfo{year}{2001}).

\bibitem[{\citenamefont{Schmidt et~al.}(2005)\citenamefont{Schmidt, Weber,
  Elmers, Forster, Reinert, H\"ufner, Escher, Merkel, Kr\"omker, and
  Funnemann}}]{schmidt05}
\bibinfo{author}{\bibfnamefont{S.}~\bibnamefont{Schmidt}},
  \bibinfo{author}{\bibfnamefont{N.}~\bibnamefont{Weber}},
  \bibinfo{author}{\bibfnamefont{H.J.}~\bibnamefont{Elmers}},
  \bibinfo{author}{\bibfnamefont{F.}~\bibnamefont{Forster}},
  \bibinfo{author}{\bibfnamefont{F.}~\bibnamefont{Reinert}},
  \bibinfo{author}{\bibfnamefont{S.}~\bibnamefont{H\"ufner}},
  \bibinfo{author}{\bibfnamefont{M.}~\bibnamefont{Escher}},
  \bibinfo{author}{\bibfnamefont{M.}~\bibnamefont{Merkel}},
  \bibinfo{author}{\bibfnamefont{B.}~\bibnamefont{Kr\"omker}},
  \bibnamefont{and}
  \bibinfo{author}{\bibfnamefont{D.}~\bibnamefont{Funnemann}},
  \bibinfo{journal}{Phys. Rev. B} \textbf{\bibinfo{volume}{72}},
  \bibinfo{pages}{064429} (\bibinfo{year}{2005}).

\bibitem[{\citenamefont{H{\"u}fner}(1994)}]{huefner}
\bibinfo{author}{\bibfnamefont{S.}~\bibnamefont{H{\"u}fner}},
  \emph{\bibinfo{title}{Photoelectron Spectroscopy}}
  (\bibinfo{publisher}{Springer-Verlag},
  \bibinfo{address}{Berlin--Heidelberg--New York}, \bibinfo{year}{1994}).

\bibitem[{\citenamefont{Grube et~al.}(1999)\citenamefont{Grube, Fietz, Tutsch,
  Stockert, and v.~L\"ohneysen}}]{grube99}
\bibinfo{author}{\bibfnamefont{K.}~\bibnamefont{Grube}},
  \bibinfo{author}{\bibfnamefont{W.~H.} \bibnamefont{Fietz}},
  \bibinfo{author}{\bibfnamefont{U.}~\bibnamefont{Tutsch}},
  \bibinfo{author}{\bibfnamefont{O.}~\bibnamefont{Stockert}}, \bibnamefont{and}
  \bibinfo{author}{\bibfnamefont{H.}~\bibnamefont{v.L\"ohneysen}},
  \bibinfo{journal}{Phys. Rev. B} \textbf{\bibinfo{volume}{60}},
  \bibinfo{pages}{11947} (\bibinfo{year}{1999}).

\bibitem[{\citenamefont{Gunnarsson and Sch{\"o}nhammer}(1983)}]{gunnarsson83}
\bibinfo{author}{\bibfnamefont{O.}~\bibnamefont{Gunnarsson}} \bibnamefont{and}
  \bibinfo{author}{\bibfnamefont{K.}~\bibnamefont{Sch{\"o}nhammer}},
  \bibinfo{journal}{Phys. Rev. B} \textbf{\bibinfo{volume}{28}},
  \bibinfo{pages}{4315} (\bibinfo{year}{1983}).

\bibitem[{\citenamefont{Gunnarsson et~al.}(1983)\citenamefont{Gunnarsson,
  Sch{\"o}nhammer, Fuggle, Hillebrecht, Esteva, Karnatak, and
  Hillebrand}}]{gunnarsson83b}
\bibinfo{author}{\bibfnamefont{O.}~\bibnamefont{Gunnarsson}},
  \bibinfo{author}{\bibfnamefont{K.}~\bibnamefont{Sch{\"o}nhammer}},
  \bibinfo{author}{\bibfnamefont{J.~C.} \bibnamefont{Fuggle}},
  \bibinfo{author}{\bibfnamefont{F.~U.} \bibnamefont{Hillebrecht}},
  \bibinfo{author}{\bibfnamefont{J.M.}~\bibnamefont{Esteva}},
  \bibinfo{author}{\bibfnamefont{R.~C.} \bibnamefont{Karnatak}},
  \bibnamefont{and}
  \bibinfo{author}{\bibfnamefont{B.}~\bibnamefont{Hillebrand}},
  \bibinfo{journal}{Phys. Rev. B} \textbf{\bibinfo{volume}{28}},
  \bibinfo{pages}{7330} (\bibinfo{year}{1983}).

\bibitem[{\citenamefont{Ehm et~al.}(2001)\citenamefont{Ehm, Reinert, Nicolay,
  Schmidt, H{\"u}fner, Claessen, Eyert, and Geibel}}]{ehm01}
\bibinfo{author}{\bibfnamefont{D.}~\bibnamefont{Ehm}},
  \bibinfo{author}{\bibfnamefont{F.}~\bibnamefont{Reinert}},
  \bibinfo{author}{\bibfnamefont{G.}~\bibnamefont{Nicolay}},
  \bibinfo{author}{\bibfnamefont{S.}~\bibnamefont{Schmidt}},
  \bibinfo{author}{\bibfnamefont{S.}~\bibnamefont{H{\"u}fner}},
  \bibinfo{author}{\bibfnamefont{R.}~\bibnamefont{Claessen}},
  \bibinfo{author}{\bibfnamefont{V.}~\bibnamefont{Eyert}}, \bibnamefont{and}
  \bibinfo{author}{\bibfnamefont{C.}~\bibnamefont{Geibel}},
  \bibinfo{journal}{Phys. Rev. B} \textbf{\bibinfo{volume}{64}},
  \bibinfo{pages}{235104} (\bibinfo{year}{2001}).

\bibitem[{\citenamefont{Fuggle et~al.}(1983)\citenamefont{Fuggle, Hillebrecht,
  Zolnierek, L\"asser, Freiburg, Gunnarsson, and Sch\"onhammer}}]{fuggle83}
\bibinfo{author}{\bibfnamefont{J.~C.} \bibnamefont{Fuggle}},
  \bibinfo{author}{\bibfnamefont{F.~U.} \bibnamefont{Hillebrecht}},
  \bibinfo{author}{\bibfnamefont{Z.}~\bibnamefont{Zolnierek}},
  \bibinfo{author}{\bibfnamefont{R.}~\bibnamefont{L\"asser}},
  \bibinfo{author}{\bibfnamefont{C.}~\bibnamefont{Freiburg}},
  \bibinfo{author}{\bibfnamefont{O.}~\bibnamefont{Gunnarsson}},
  \bibnamefont{and}
  \bibinfo{author}{\bibfnamefont{K.}~\bibnamefont{Sch\"onhammer}},
  \bibinfo{journal}{Phys. Rev. B} \textbf{\bibinfo{volume}{27}},
  \bibinfo{pages}{7330} (\bibinfo{year}{1983}).

\bibitem[{\citenamefont{Allen et~al.}(1986)\citenamefont{Allen, Oh, Gunnarsson,
  Sch\"onhammer, Maple, Torikachvili, and Lindau}}]{allen86}
\bibinfo{author}{\bibfnamefont{J.~W.} \bibnamefont{Allen}},
  \bibinfo{author}{\bibfnamefont{S.~J.} \bibnamefont{Oh}},
  \bibinfo{author}{\bibfnamefont{O.}~\bibnamefont{Gunnarsson}},
  \bibinfo{author}{\bibfnamefont{K.}~\bibnamefont{Sch\"onhammer}},
  \bibinfo{author}{\bibfnamefont{M.~B.} \bibnamefont{Maple}},
  \bibinfo{author}{\bibfnamefont{M.~S.} \bibnamefont{Torikachvili}},
  \bibnamefont{and} \bibinfo{author}{\bibfnamefont{I.}~\bibnamefont{Lindau}},
  \bibinfo{journal}{Adv. in Phys.} \textbf{\bibinfo{volume}{35}},
  \bibinfo{pages}{275} (\bibinfo{year}{1986}).

\bibitem[{\citenamefont{Witkowski et~al.}(1997)\citenamefont{Witkowski,
  Bertran, and Malterre}}]{witkowski97b}
\bibinfo{author}{\bibfnamefont{N.}~\bibnamefont{Witkowski}},
  \bibinfo{author}{\bibfnamefont{F.}~\bibnamefont{Bertran}}, \bibnamefont{and}
  \bibinfo{author}{\bibfnamefont{D.}~\bibnamefont{Malterre}},
  \bibinfo{journal}{Phys. Rev. B} \textbf{\bibinfo{volume}{56}},
  \bibinfo{pages}{15040} (\bibinfo{year}{1997}).

\bibitem[{\citenamefont{Witkowski et~al.}(2001)\citenamefont{Witkowski,
  Bertran, and Malterre}}]{witkowski01}
\bibinfo{author}{\bibfnamefont{N.}~\bibnamefont{Witkowski}},
  \bibinfo{author}{\bibfnamefont{F.}~\bibnamefont{Bertran}}, \bibnamefont{and}
  \bibinfo{author}{\bibfnamefont{D.}~\bibnamefont{Malterre}},
  \bibinfo{journal}{J. Electron Spectrosc. Relat. Phenom.}
  \textbf{\bibinfo{volume}{117}} (\bibinfo{year}{2001}).

\bibitem[{\citenamefont{Bickers et~al.}(1987)\citenamefont{Bickers, Cox, and
  Wilkins}}]{bickers87}
\bibinfo{author}{\bibfnamefont{N.~E.} \bibnamefont{Bickers}},
  \bibinfo{author}{\bibfnamefont{D.~L.} \bibnamefont{Cox}}, \bibnamefont{and}
  \bibinfo{author}{\bibfnamefont{J.~W.} \bibnamefont{Wilkins}},
  \bibinfo{journal}{Phys. Rev. B} \textbf{\bibinfo{volume}{36}},
  \bibinfo{pages}{2036} (\bibinfo{year}{1987}).

\end{thebibliography}
\end{document}